\documentclass{PoS}

\usepackage{braket}

\title{Dust in Protoplanetary Disks: A Clue as to the Critical Mass of Planetary Cores}

\ShortTitle{Dust in Protoplanetary Disks: A Clue as to the Critical Mass of Planetary Cores}

\author{\speaker{Yasuhiro Hasegawa}%
         \thanks{EACOA follow}\\
        ASIAA\\
        E-mail: \email{yasu@asiaa.sinica.edu.tw}}

\author{Ralph E Pudritz\\
        Department of Physics and Astronomy, McMaster University\\
        Origins Institute, McMaster University\\
        E-mail: \email{pudritz@physics.mcmaster.ca}}

\abstract{
Dust in protoplanetary disks is widely recognized as the building blocks of planets that are eventually formed in the disks. 
In the core accretion scenario, one of the standard theories of gas giant formation, 
the abundance of dust in disks (or metallicity, [Fe/H]) plays a crucial role in regulating the formation of cores of gas giants 
that proceeds via collisions of dust and planetesimals in disks.
We present our recent progress on the relationship between the metallicity and planet formation, 
wherein planet formation frequencies (PFFs) as well as the critical mass of planetary cores ($M_{c,crit}$) that can initiate gas accretion 
are statistically examined.
We focus on three different planetary populations that are prominent in the distribution of observed exoplanets in the mass-semimajor axis diagram: 
hot Jupiters, exo-Jupiters that are densely populated around 1 AU, and low-mass planets in tight orbits, also known as super-Earths and hot Neptunes.
We show that the resultant PFFs for both Jovian planets are correlated positively with the metallicity of disks whereas low-mass planets 
form efficiently for a wide range of metallicities ($-0.6\le$[Fe/H]$\le 0.6$). 
This is consistent with the so-called Planet-Metallicity correlation that is inferred from both the radial velocity and transit observations. 
By plotting the statistically averaged value of $M_{c,crit}$ (defined as $\braket{M_{c,crit}}$) as a function of metallicity, 
we find that the correlation originates from the behavior of $\braket{M_{c,crit}}$ that increases steadily with metallicity for two kinds of the Jovian planets 
while the low-mass planets obtain a rather constant value for $\braket{M_{c,crit}}$.
Such a different behavior in $\braket{M_{c,crit}}$ enables one to define transition metallicities (TMs) above which 
the Jovian planets gain a larger value of $\braket{M_{c,crit}}$ than the low-mass planets, 
and hence gas giant formation takes place more efficiently.
We find that TMs locate at [Fe/H]$\simeq -0.2$ to $-0.4$, and are sensitive to the important parameter that involves $M_{c,crit}$. 
We demonstrate, by comparing with the observations, that a most likely value of $M_{c,crit}$ is $\simeq 5M_{\oplus}$, 
which is smaller than the widely adopted value in the literature ($\simeq 10M_{\oplus}$).  
Our results therefore suggest that opacities in the atmospheres surrounding planetary cores play an important role for lowering $M_{c,crit}$.
}

\FullConference{The Life Cycle of Dust in the Universe: Observations, Theory, and Laboratory Experiments\\
                 18-22 November, 2013\\
                 Taipei, Taiwan}

\begin{document}

\section{Introduction}

The dust cycle in the Universe involves a number of interesting physical processes. 
One of the fundamental components in the cycle is the formation of planets in protoplanetary disks 
that can provide potential abodes for developing life. 

The rapid accumulation of observed exoplanets, 
that occurred since the first discovery of the massive planet around the solar-type star \cite{mq95}, 
has opened up a new window to understand planet formation in protoplanetary disks \cite{us07}. 
One of the important relations that are derived  from the observations is the so-called Planet-Metallicity correlation: 
more massive planets are observed preferentially for stars with higher metallicities 
while low-mass planets are detected for a wide range of stellar metallicities ($-0.6\le$[Fe/H]$\le 0.6$).

The correlation is considered as strong evidence that most observed gas giants at $r \le 10$ AU are formed via the core accretion scenario, 
wherein gas giants undergo sequential accretion of dust and gas \cite{p96}. 
Population synthesis models based on this scenario indeed confirm that the resultant planetary populations reproduce the correlation well \cite{il04ii}.
In the picture, the most critical process, that results in the observed Planet-Metallicity correlation, arises in the phase of core formation. 
Then, an intriguing question can be posed: can the core accretion picture provide some kind of correlation between the core mass and the metallicity, 
which can act as a basis for the observed Planet-Metallicity correlation?

\section{Evolutionary Tracks of Planets Forming at Planet Traps}

We explore the problem, by adopting a formulation developed in a series of our papers \cite{hp11,hp12,hp13a,hp14a}. 
The model is built for computing the evolution of both forming and migrating planets in viscously evolving gas disks with photoevaporation of gas 
in the mass-semimajot axis diagram. 
Since the detailed description of theoretically computed evolutionary tracks of planets can be found somewhere \cite{hp12,hp14a}, 
we focus on a couple of the important processes in the model (see the left panel of Figure \ref{fig1}).

The key components of the model are planet traps that serve as barriers for rapid type I migration for planetary cores \cite{hp11}. 
The presence of planet traps in protoplanetary disks arises naturally from 
both local variations in disk structures from simple power-law approximations as well as the high sensitivity of the migration to them.
It is therefore expected that planet formation proceeds locally at planet traps, because cores of gas giants are captured and accumulate there.
Thus, planet traps provide a good connection between the disk structure and planet formation. 

We discuss the part of planetary growth in our model. 
The primary focus of this work is on the effect of the core mass on the observed Planet-Metallicity correlation. 
In order to proceed, 
we examine the critical mass of planetary cores ($M_{c,crit}$) that can initiate gas accretion onto the cores.
It is a key quantity because the efficiency of the subsequent gas accretion is regulated largely by the value of $M_{c,crit}$, 
which is written as \cite{il04ii,hp12}
\begin{equation}
 M_{c,crit} \simeq M_{c,crit0} \left( \frac{\dot{M}_c}{10^{-6}M_{\oplus} \mbox{ yr}^{-1}} \right)^{1/4},
 \label{m_ccrit}
\end{equation}
where $\dot{M}_c$ is the accretion rate of planetesimals by cores, and $M_{c,crit0}$ is defined as 
\begin{equation}
\label{m_ccrit0}
M_{c,crit0} \equiv 10 M_{\oplus} \left( \frac{\kappa} {1 \mbox{ cm}^2 \mbox{ g}^{-1}} \right)^{0.2-0.3},
\end{equation} 
where $\kappa$ is the grain opacity in the envelopes surrounding cores \cite{ine00}. 
We emphasize that $M_{c,crit0}$ is a parameter in our formalism. 
Thus, we will perform a parameter study for the value of $M_{c,crit0}$ below and 
discuss how important the effect of $\kappa$ is for decreasing/increasing $M_{c,crit0}$.

Figure \ref{fig1} (left) shows how the value of $M_{c,crit}$ is determined along the tracks (see the circles).
These examples show different values of $M_{c,crit}$ for different planetary populations.
Note that all the planetary populations are generated via the same mechanism in our model. 
As a result, some tracks, that end up in the zone of the low-mass planets, 
are considered to form failed cores of gas giants and/or mini-gas giants (see the dotted line). 

\section{The Resultant PFFs and Critical Mass of Planetary Cores}

Making use of the above framework in which a complete set of evolutionary tracks of planets forming at planet traps are theoretically computed, 
we will examine how the value of $M_{c,crit}$ changes with the disk metallicity 
that links directly to the efficiency of forming cores of gas giants.
In order to proceed, we further developed a new statistical approach \cite{hp13a}.

\begin{figure*}
\begin{minipage}{15cm}
\begin{center}
\includegraphics[width=7cm]{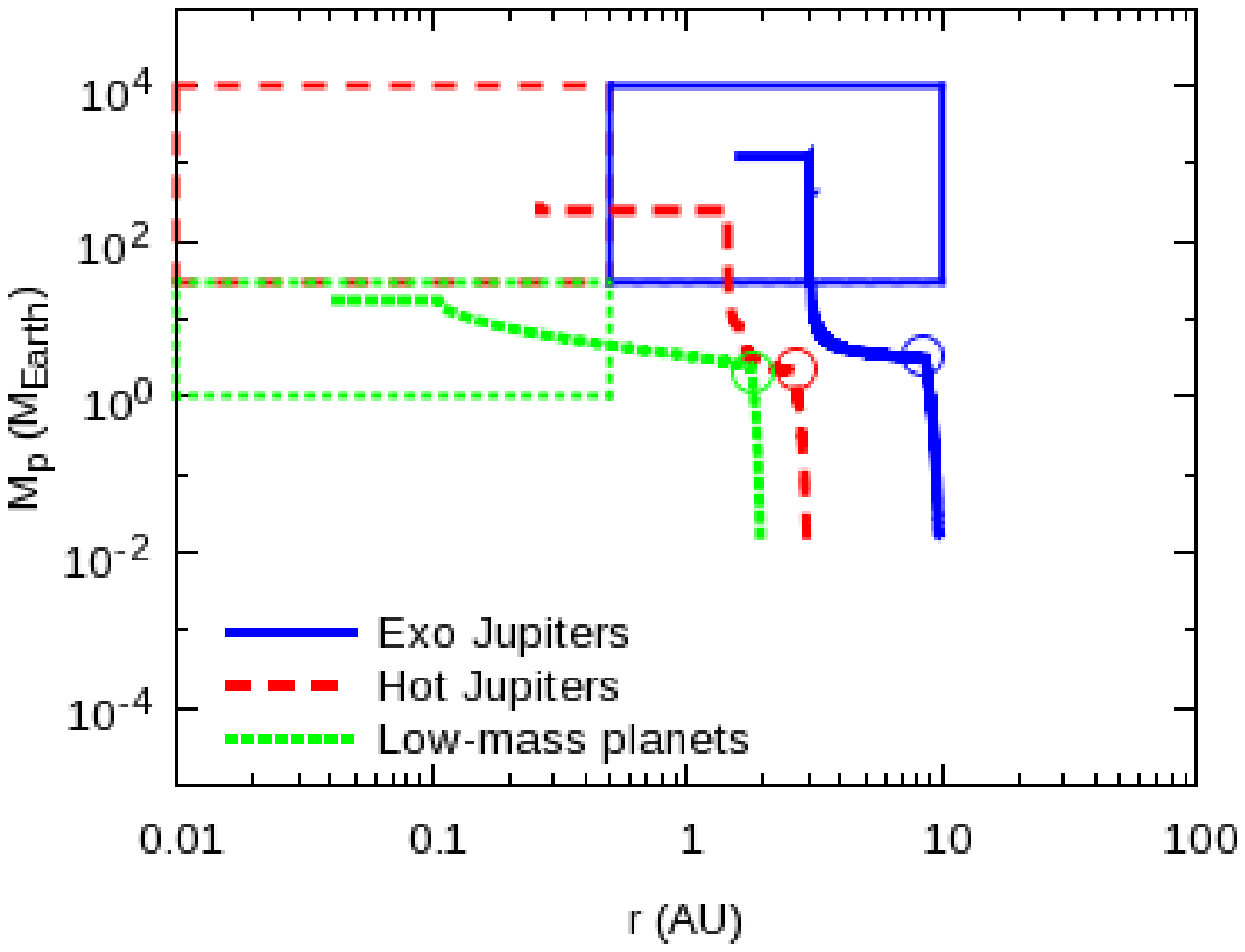}
\includegraphics[width=7cm]{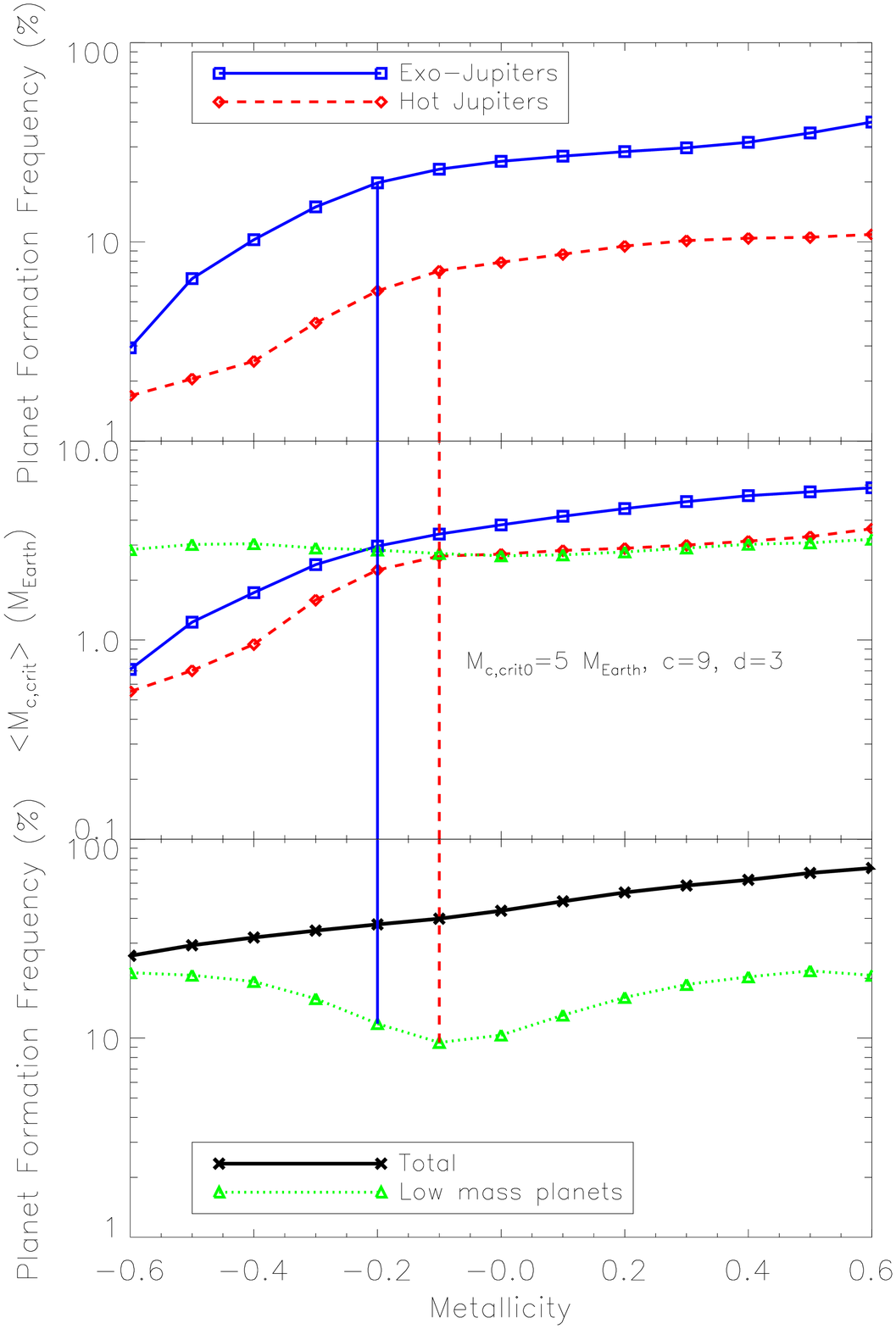}
\caption{
{\it Left:} Examples of theoretically computed evolutionary tracks for planets forming at planet traps.
The value of $M_{c,crit}$ is denoted by the circle.
{\it Right:} The resultant PFFs as well as statistically averaged value of $M_{c,crit}$ (denoted by $\braket{M_{c,crit}}$), as a function of metallicity.
The top panel shows the PFFs for the hot and exo-Jupiters whereas the bottom is for the total and the low-mass planets.
The middle panel shows the behavior of $\braket{M_{c,crit}}$ for the three different planetary populations.
For our fiducial case ($M_{c,crit0}=5M_{\oplus}$), the TM for the hot and exo-Jupiters, that is denoted by the vertical line, 
is [Fe/H]$=-0.1$ and $-0.2$, respectively.
}
\label{fig1}
\end{center}
\end{minipage}
\end{figure*}

The approach consists of three procedures. 
In the first one, we divide the mass-semimajor axis diagram into three zones (see Figure \ref{fig1}). 
Zoning the diagram is motivated by the exoplanet observations which show that 
most observed planets ($r \le 10$ AU) distribute in these zones \cite{hp13a}. 
Then, we compute a large number of tracks ($N$), by varying both the disk accretion rate and the disk lifetime parameters. 
In addition to the stellar parameters, these three sets of parameters cover the full parameter space 
that regulates planet formation in protoplanetary disks.
Based on our preliminary results, $N=300$ is good enough to get results to converge.
Finally, we count the end-points of tracks that fill out each zones. 
By integrating such a number with weight functions of both the disk accretion rate as well as the disk lifetime, 
we can estimate planet formation frequencies (PFFs) for each zones. 
The resultant PFFs are comparable to the results of the standard population synthesis calculations, 
because we adopt an analytical modeling for both the two weight functions that can reproduce the disk observations. 
For the value of $M_{c,crit}$, we adopt the same approach as above for deriving the statically averaged value ($\braket{M_{c,crit}}$).

In summary, we can compare our results with the exoplanet observations directly without the standard population synthesis calculations being performed.

Figure \ref{fig1} (right) shows the resultant PFFs and $\braket{M_{c,crit}}$ as a function of metallicity.
We first discuss the PFFs for the hot and exo-Jupiters (see the dashed and the solid line, respectively on the top panel). 
Our results show that both the Jovian planets attain a larger value of PFFs for higher metallicities 
whereas their PFFs decrease rapidly toward lower metallicities.
In addition, we find that the PFFs for the exo-Jupiters are always higher than those for the hot Jupiters for the metallicity range 
in which most observed exoplanets are confined ($-0.6\le$[Fe/H]$\le 0.6$).

The sensitivity of the PFFs for both the Jovian planets to the metallicity arises as a direct reflection of the core accretion scenario 
in which the efficiency of forming gas giants is an increasing function of metallicity. 
This is consistent with the observed metallicity correlation for gas giants.
The formation of the exo-Jupiters with higher PFFs than the hot Jupiters for a wide range of metallicities is regarded as 
the generalization of \cite{hp13a} which shows that the zone of the exo-Jupiters is the most preferred place for gas giants to end up. 
This is also in good agreement with the exoplanet observations in which most massive planets distribute around $r=1$ AU with 
the presence of fewer hot Jupiters.

We then discuss the results of the PFFs for the total as well as the low-mass planets (see the thick and the dotted line, respectively on the bottom panel).
The results show that the total PFFs increase steadily with increasing the metallicity.
This behavior is again understood by the nature of the core accretion picture. 
The PFFs for the low-mass planets, on the contrary, are quite insensitive to the metallicity, 
although one may notice some oscillation in their behavior as a function of metallicity.
As already pointed out, they are built like forming gas giants: only the difference between them is the efficiency of gas accretion onto the cores. 
Thus, our results indicate that a large fraction of observed super-Earths and hot Neptunes may be formed as failed cores of gas giants and/or mini-gas giants.

As a conclusion, our resultant PFFs indicate that 
the combination of planet traps with the core accretion scenario provides a good physical explanation for the observations.

We finally discuss the behavior of $\braket{M_{c,crit}}$ for the three different planetary populations as a function of metallicity.
Figure \ref{fig1} (right, middle) shows that $\braket{M_{c,crit}}$ for both the Jovian planets is an increasing function of metallicity 
whereas that for the low-mass planets is almost constant with varying the metallicity.
The results also show that $\braket{M_{c,crit}}$ for the exo-Jupiters has a larger value than the hot Jupiters for the metallicity range under consideration.

It is important to note that the resultant PFFs for both the Jovian planets exactly follow their behaviors of $\braket{M_{c,crit}}$: 
their PFFs become higher when $\braket{M_{c,crit}}$ achieves a larger value, and vice versa. 
This is one of the most important findings in this work.
Thus, the observed Planet-Metallicity correlation for gas giants originates from the core mass-metallicity correlation, 
wherein more massive cores of gas giants are formed for the higher metallicity environment.
Furthermore, a larger value of $\braket{M_{c,crit}}$ for the exo-Jupiters than the hot Jupiters provides an additional explanation of 
why the zone of the exo-Jupiters is the most preferred place for gas giants. 
Our results indicate that, when planet formation proceeds predominately at planet traps, 
planets that eventually fill out the zone of the exo-Jupiters undergo very efficient formation of planetary cores. 
As a result, more massive planets tend to distribute in the zone of the exo-Jupiters than in that of the hot Jupiters.

The behavior of $\braket{M_{c,crit}}$ for the low-mass planets is additional evidence that 
low-mass planets formed in our model are insensitive to the metallicity.

The comparison of $\braket{M_{c,crit}}$ between the Jovian and the low-mass planets provides another important insight as to the theory of planet formation. 
Since low-mass planets are built as failed cores of gas giants and/or mini-gas giants in our model,  
the value of $\braket{M_{c,crit}}$ for such planets can be used as a threshold value for forming gas giants. 
Specifically, it is required that the core mass of a protoplanet should be larger than $\braket{M_{c,crit}}$ of the low-mass planets, 
in order for the protoplanet to undergo efficient gas accretion and to grow up to a gas giant.
We can therefore estimate transition metallicities (TMs) above which gas giant formation proceeds efficiently, 
and hence the population of gas giants dominates over the low-mass planets.
Our fiducial case shows that the TM for the hot and exo-Jupiters is [Fe/H]$=-0.1$ and [Fe/H]$=-0.2$, respectively 
(see the vertical dashed and solid line, respectively).

\section{Comparison with Observations}

As demonstrated above, it is useful to plot $\braket{M_{c,crit}}$ for the three different planetary populations as a function of metallicity. 
Examination of the diagram provides not only the origin of the observed Planet-Metallicity correlation, but also the estimate of TMs. 
We here examine how the behavior of TMs links to the theory of planet formation.
To achieve the goal, we perform a parameter study on the value of $M_{c,crit0}$ (see equations (\ref{m_ccrit}) and (\ref{m_ccrit0})). 
Note that the other parameters, $c$ and $d$, involve the gas accretion (see \cite{hp14a} for the complete discussion). 

\begin{figure*}
\begin{minipage}{15cm}
\begin{center}
\includegraphics[width=7cm]{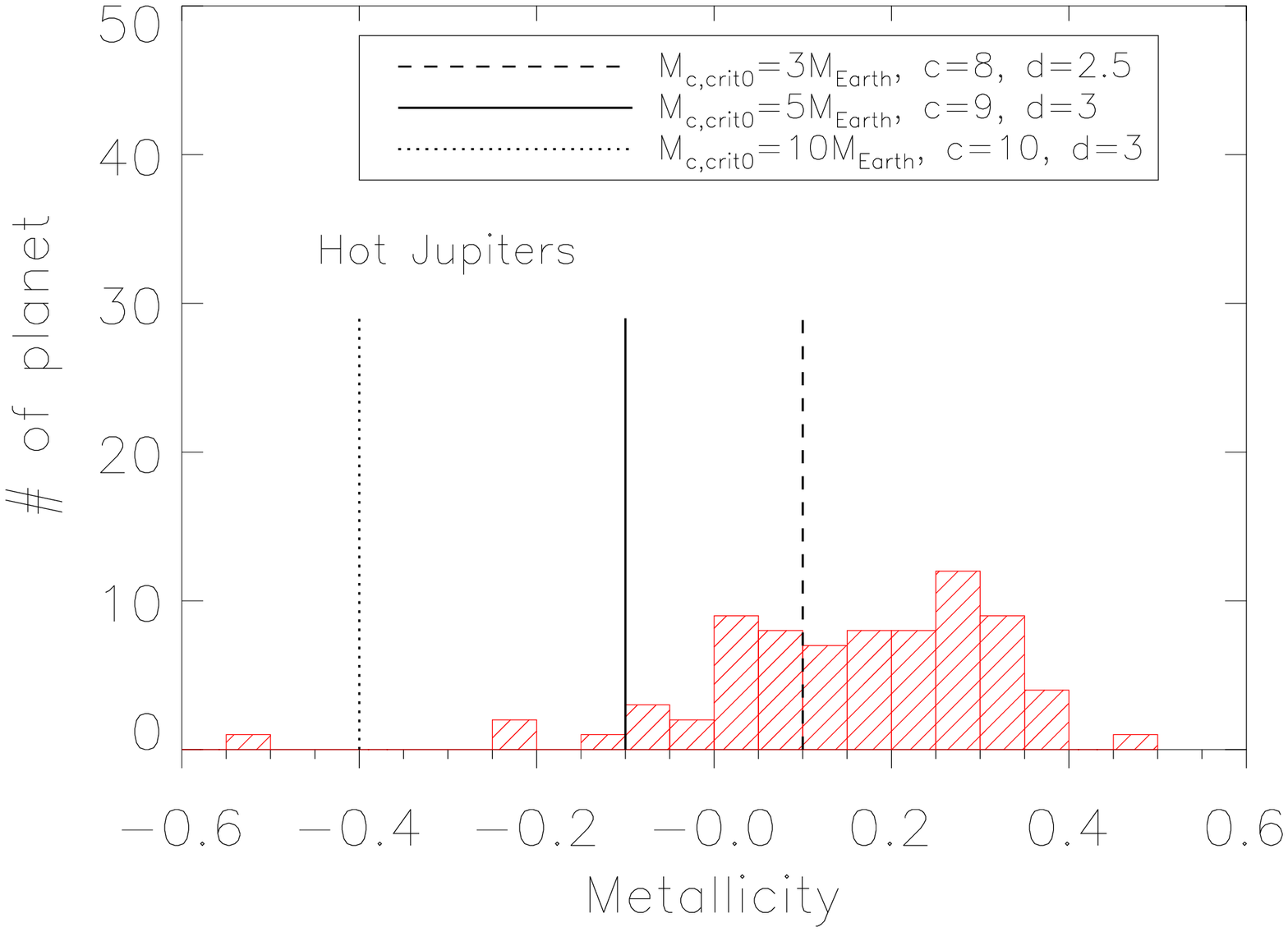}
\includegraphics[width=7cm]{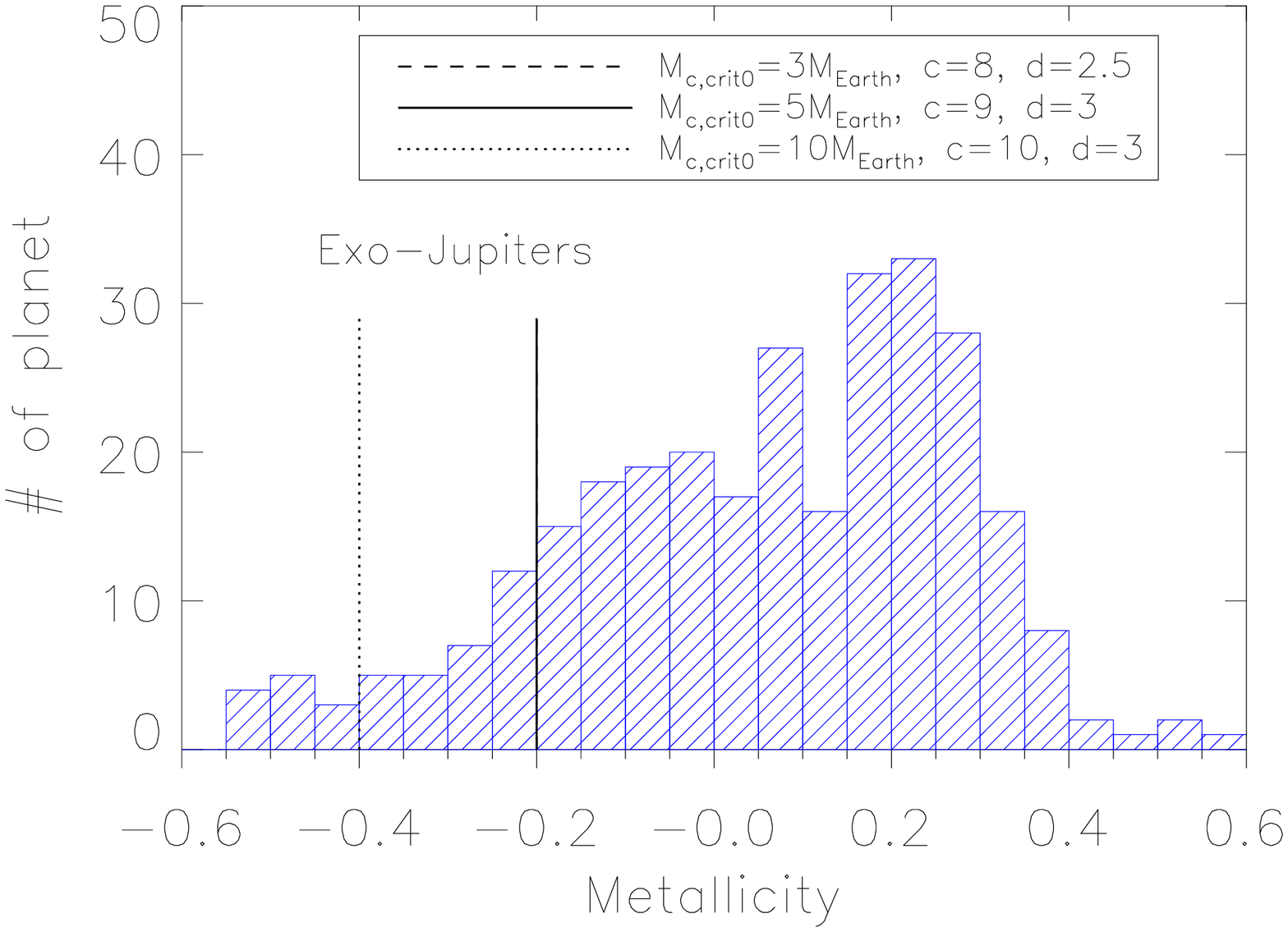}
\caption{The comparison of our results with the radial velocity observations. 
The number of the observed hot and exo-Jupiters is plotted in the histogram. 
The values of TMs for both the hot and exo-Jupiters are also shown (see the vertical lines). 
Note that the dashed line on the right panel completely matches the solid line. 
Our results imply that the currently available data of exoplanets are likely to be fitted well by $M_{c,crit0}(=5M_{\oplus})$ 
that is smaller than $10M_{\oplus}$.}
\label{fig2}
\end{center}
\end{minipage}
\end{figure*}

Figure \ref{fig2} shows the position of TMs for three different values of $M_{c,crit0}$ (see the vertical lines). 
In order to compare with the observations, 
the number of observed hot and exo-Jupiters is plotted as a function of metallicity on the left and the right panel, respectively.
Our results indicate that the case of $M_{c,crit0}=5M_{\oplus}$ provides the best fit to the currently available observational data, 
assuming the data is roughly complete over the metallicity range ($-0.6\le$[Fe/H]$\le 0.6$). 
Thus, we can conclude that 
the widely adopted value, $M_{c,crit0}=10M_{\oplus}$, in the literature is very likely to be too large to reproduce the exoplanet observations.

What is a physical origin to lower the value of $M_{c,crit0}$?
This can be caused by the grain opacity in the atmosphere around the cores of gas giants.
Adopting the relationship between the opacity and $M_{c,crit0}$ (see equation (\ref{m_ccrit0})), 
our results show that the observed massive planets have the opacity of the atmosphere that is roughly one order of magnitude lower than 
the canonical value: $\kappa \simeq 0.1\mbox{ cm}^2 \mbox{ g}^{-1}$.
This lower value is still consistent with more detailed calculations \cite{ine00,hi10}.

In summary, we have succeeded well in deriving invaluable constraints on the theory of gas giant formation  
by developing our version of population synthesis calculations and comparing our results with the exoplanet observations. 
It is obviously important to undertake a more intensive survey of exoplanets orbiting around  metal-poor stars ($-0.4\le$[Fe/H]$\le -0.1$) 
for better understanding planet formation in protoplanetary disks.
 
\end{document}